\documentclass{article}

\usepackage[a4paper, left=20mm, top=20mm,]{geometry}
\usepackage[utf8]{inputenc}
\usepackage{hyperref}
\hypersetup{colorlinks = true, citecolor = blue, linkcolor = blue}

\usepackage{mathtools}
\usepackage{amsfonts}
\usepackage{csquotes}
\usepackage{booktabs}
\usepackage{graphicx}
\usepackage{xfrac}
\usepackage{soul}
\usepackage{authblk}

\DeclarePairedDelimiter\autobracket{(}{)}
\DeclarePairedDelimiter\abs{\lvert}{\rvert}
\newcommand{\br}[1]{\autobracket*{#1}}

\newcommand{\Prob}{\text{Pr}}
\newcommand{\R}{\rho}
\newcommand{\T}{\theta}
\newcommand{\TT}{\tilde{\theta}}

\undef\S
\newcommand{\S}{\sigma}

\usepackage{natbib}
\title{Approximate Bayes factors for unit root testing}

\author[1]{Martin Magris\thanks{Corresponding author, magris@ece.au.dk. Submitted for consideration at the 2021 annual conference of the {\it International Association for Applied Econometrics} (IAAE). }}
\author[1]{Alexandros Iosifidis}
\affil[1]{\small \it Department of Electrical and Computer Engineering, Aarhus University, Denmark}
\begin{document}
\date{}

\newpage

\maketitle
\begin{abstract}
This paper introduces a feasible and practical Bayesian method for unit root testing in financial time series. We propose a convenient approximation of the Bayes factor in terms of the Bayesian Information Criterion as a straightforward and effective strategy for testing the unit root hypothesis. Our approximate approach relies on few assumptions, is of general applicability, and preserves a satisfactory error rate. Among its advantages, it does not require the prior distribution on model's parameters to be specified. Our simulation study and empirical application on real exchange rates show great accordance between the suggested simple approach and both Bayesian and non-Bayesian alternatives.\\

\noindent {\it Keywords}: Unit root inference; Bayesian analysis; Bayes factor; BIC\\
{\it JEL classification:} C11; C12; C22
\end{abstract}

\section{Introduction}
In time series analysis the interest is often over some persistence properties of the process in the medium-long term. 
In this regard, unit roots are important since they relate to persistence of shocks and non-mean-reverting dynamics \citep[e.g.][]{campbell1991pitfalls}. In other words, unit roots go in hand with non-stationarity. This has severe implications in the applicability of standard econometrics techniques, among which spurious regression \citep{granger1974spurious,phillips1986understanding} is the most notable one.
Non-stationarity appears to arise quite commonly in economic time series, and with little surprise unit root testing has been one of the most proficient research areas in econometrics. Most theoretical and empirical work in the domain of non-stationary time series relies on classical frequentist methods, which stand as the reference approach.

Among others, \citep{zellner1971introduction,geweke1988secular,poirier1988frequentist} showed that the Bayesian framework is in general well-suited to inferential problems in econometrics, and it was since \citep{sims1988bayesian} that a number of Bayesian methods for unit root testing and the corresponding Bayesian unit root literature developed.
The earliest works appear remarkably optimistic and confident about the impact that a Bayesian approach could have in unit root testing, even claiming an overall superiority of the Bayesian approach over classical methods \citep{sims1991understanding}. On the contrary, the adoption and development of Bayesian methods would shortly appear to be much less straightforward and very debated. It was in \citep{phillips1991criticize} that a reconciliation with frequentist methods was first discussed under the light of impartial and objective Bayesian methods. This work raised several further issues related to the Bayesian approach in unit root testing that stimulated active research and strong debates.
\citep[][]{phillips1991rebus,phillips1991criticize} advanced the idea that the achievement of impartial Bayesian analysis through flat priors on the parameters is not well-suited in time series models. Actually, flat priors over the autoregressive parameter achieve the opposite effect and have shown to be quite informative \citep[e.g.][among the others]{kim1991flat,schotman1991bayesian,phillips1991criticize,leamer1991comment}. A long debate on appropriate uninformative priors for Bayesian unit root inference followed (see Section \ref{sec:literature}). The determination of suitable input information through the prior distribution, which is generally the major reason for the divergence between the classical and Bayesian approach, is tightly bonded to the exact hypothesis being tested. 
For a simple autoregressive process of order one, hereafter denoted by AR(1), the unit root inference problem  corresponds to testing the null hypothesis of the autoregressive parameter being equal to one. In general, we shall emphasize the exact/point-wise/non-interval nature of an hypothesis by referring to it as {\it point} hypothesis.
The goal of testing a point null hypothesis cannot be easily achieved with continuous priors, as the consequent continuous posterior would assign zero weight to the unit root hypothesis. 
Feasible tests can either use discontinuous priors that assign a non-zero mass to the unit root hypothesis and distribute the reaming one over some interval \citep[e.g.][]{schotman1991exchange,dejong1991reconsidering}, or test closely-related non-point hypotheses with continuous priors \citep[][is instructive as it considers three possible nulls]{koop1994objective}. The first approach is susceptible to poor objectivity \citep{phillips1993long} while the second one does not properly match the exact and exclusive purpose of testing the unit root hypothesis \citep{schotman1991bayesian}. 
Consequently, the Bayesian analysis can be respectively based on two radically different approaches: on Bayes factors and odds ratios, or Bayes confidence sets and probability intervals over the posterior, with the first one being of difficult interpretation in terms of traditional p-values \citep{berger1987testing}.
Furthermore, the inclusion of an intercept, a trend, or any richer structure in a simple AR(1) model, are not smooth extensions, as prior beliefs on the autoregressive parameter generally change according to the particular deterministic component added to the model \citep{schotman1991bayesian}. 
Also, the particular form under which a model is expressed, e.g. \enquote{structural} or \enquote{reduced} from, plays a role in unveiling feasible directions for the Bayesian analysis. Not less importantly,  \citep{schotman1991bayesian,uhlig1994macroeconomists,lubrano1995testing} underline the importance of the conditioning set, and in particular the sensitivity of the whole inferential procedure on the first observation.
Lastly, Bayesian methods are generally known for not being keen on simple algebra: also in unit root inference numerical methods are often required for achieving approximate solutions \citep[e.g.][]{zivot1994bayesian}.

In this paper, we propose an approximate Bayesian unit root testing procedure that mitigates the above-mentioned criticalities, especially the choice of the prior. In particular, we focus on the simple AR(1) dynamics, which is a fundamental process and a major ingredient in the theory of non-stationary time series and unit root testing, and on testing the point null unit root hypothesis. We apply standard approximation results to obtain a generic Bayesian testing procedure based on approximate Bayes factors and thus approximate posterior odds. 
With an asymptotic error rate sufficient to guarantee the applicability of the proposed method also for samples of moderate size, the approximate form of the Bayes factor is independent on the choice of the priors, it is remarkably simple to compute, and scales to more complex models as long as their maximum likelihood estimates are attainable. Indeed, our approximate Bayes factor is formulated as a simple function of the well-known Bayesian Information Criteria (BIC), being thus easy to implement and attractive for empirical research.
Although the BIC approximation of Bayes factors has proved to be a valid tool in different fields, its use in unit root testing has not been investigated, though it appears to be very well-suited for this class of problems.
In the empirical sections, we propose a Monte Carlo experiment and an empirical application on real exchange rates, and analyze the performance of our BIC approach with respect to other competing Bayesian methods and the most widespread frequentist Dickey-Fuller test \citep{dickey1981likelihood}. Our experiment validates the proposed approach, as it stands out as a feasible and viable simple alternative for Bayesian unit root inference.

This paper is organized as follows. Section \ref{sec:literature} reviews the literature on approaches, issues and advances in Bayesian unit root testing, underlying the numerous problems that arise in this context. Section \ref{sec:evidence} introduces Bayes factors and generally discusses the Bayesian testing framework based on {\it evidence}. Section \ref{sec:approximations} introduces our suggested testing approach based on the BIC approximation of Bayes factors. Section \ref{sec:simulation} reports the results of the Monte Carlo study, while an empirical application on real exchange rates is presented in Section \ref{sec:empirical}. Section \ref{sec:conclusion} concludes and suggests directions for future research. The Appendix collects details on some benchmark Bayesian unit root tests and further results from the simulation study.

\section{Bayesian unit root testing}  \label{sec:literature}

On the premise that the asymptotic distribution theory changes discontinuously between the stationary and unit root case, with the classical hypothesis testing appearing as a not reasonable inferential procedure opposed to a Bayesian flat prior, the Bayesian analysis of unit root models was first suggested in \citep{sims1988bayesian}.
Since then, a wide and rich literature on the field pinpointed the advantages and flaws of the Bayesian approach making it complex and long-debated.
First and most notably, the identification of a suitable prior
is in this context remarkably difficult and widely discussed. Besides this, there are several issues associated with model specification and formulation, the role played by the initial observation, issues related to the invariance of the prior under different sampling frequencies, and computational arguments. In the following, we review the most relevant aspects of these issues. For a more comprehensive overview on the topic see e.g. \citep[][Ch. 8]{maddala1998unit}, and the articles in following dedicated special issues: {\it Journal of Applied Econometrics} (1991, vol. 6, n.4), {\it Econometric Theory} (1994, vol. 10) and {\it Journal of Econometrics} (1995, vol. 69, n.1).

\subsection{The choice of the prior}
Here we shortly outline the setting based on flat priors adopted, among the others, by \citep[][]{sims1991understanding,sims1988bayesian,geweke1988secular,thornber1967finite,zellner1971introduction,schotman1991bayesian}.
Consider the simple AR(1) model:
$$
x_t = \rho x_{t-1}+u_t \text{,}
$$
and assume a flat prior for $\br{\rho,\sigma}$, $\pi\br{\rho,\sigma} \propto 1/\sigma$ with $-1<\rho<1$, and $\sigma>0$ being the standard deviation of the normal i.i.d. error  $u$.
$\rho$ is the autoregressive parameter and $\rho = 1$ corresponds to the unit root hypothesis of interest, under which the AR(1) model reduces to a Brownian motion.
Be $x_0$ the initial starting value of $T$ consecutive observations. With Gaussian likelihood
$$
L\br{x|\rho,\sigma,x_0} = \br{2\pi}^{-T/2} \sigma^{-T} \exp\br{-\frac{\sum_{t=1}^T\br{x_t-\rho x_{t-1}}^2}{2\sigma^2}}
$$
the joint posterior for $\br{\rho,\sigma}$ is given by
$$
\pi\br{\rho,\sigma|x,x_0} \propto \sigma^{-T-1} \exp\br{-\frac{\sum_{t=1}^T\br{x_t-\rho x_{t-1}}^2}{2\sigma^2}} = \sigma^{-T-1} \exp \left[ -\frac{R+\br{\rho-\hat{\rho}}^2Q}{2\sigma^{2}} \right] \, \text{,}
$$
with $\hat{\R} = \sum x_t x_{t-1}/\sum x_{t-1}^2$ being the OLS estimate for $\R$, $R=\sum \hat{\epsilon}^2_t = \sum \br{x_t-\R \hat{x_t}}^2$ the residual sum of squares, and $Q = \sum x_{t-1}^2$.
For the above joint posterior, one obtains the following margins:
\begin{align*}
    \pi\br{\rho|x,x_0} &\propto {R+\br{\rho-\hat{\rho}}^2Q}^{-T/2}\\ 
    \pi\br{\sigma|x,x_0} &\propto \sigma^{-T}\exp\br{-\frac{R}{2\sigma^2}} \text{.} \nonumber
\end{align*}
Our notation distinguishes between priors and posteriors depending on the conditioning set: $\pi\br{\cdot}$ opposed to $\pi\br{\cdot|\text{data}}$, respectively.

The marginal posterior for $\rho$ has the form of a symmetric univariate t-distribution, centered around the OLS estimate $\hat{\rho}$, while the marginal posterior for $\sigma$ is an inverted gamma-2 distribution \citep{zellner1971introduction}. \cite{sims1991understanding} conclude that classical methods relying on asymmetric distributions of the OLS estimator of $\rho$, such as the Dickey-Fuller statistics, attribute too much weight to large values of $\rho$, while the above Bayesian framework based on flat prior is a more logical and sounder basis for inference than classical testing. 
This argument is advanced by comparing the distributions $\rho|\hat{\rho} = 1$ and $\hat{\rho}|\rho = 1$, with the first being the posterior distribution of the true parameter with the estimated parameter taken as given (Bayesian approach), and the second being the sampling distribution of the estimated parameter under the value of the true parameter (classical approach). 
While classical methods are generally based on an asymmetric and nonstandard distribution for the autoregressive parameter, Bayesian methods lead to a symmetric and standard posterior. The asymmetry in $\hat{\rho}|\rho = 1$ drives the argument that classical procedures based on p-values are misleading. 

A similar approach is that developed in \citep{schotman1991bayesian}, see Appendix \ref{sec:appendix_svd} for a detailed description. This appears to be the most widespread setting for Bayesian unit root testing, commonly referenced also in the recent literature and here used as a benchmark. It takes $\rho \in  \{ S,1 \}$, $S=\{ \rho|-1 < a \leq \rho < 1 \}$, and specifies the priors for $\rho$ and $\sigma$ as
\begin{equation*}
    \Pr\br{\rho=1} = \pi_0 \, \text{,} \qquad 
    \Prob\br{\rho|\rho \in S} = \frac{1}{1-a} \, \text{,} \qquad 
    \Prob\br{\sigma} \propto \frac{1}{\sigma} \, \text{.}
\end{equation*}
That is, $\rho$ is taken uniform over $S$ and with probability mass $\pi_0$ on $\rho = 1$, $\sigma $ and $\rho$ are independent. The mass at $\rho = 1$ is intended to allow for a feasible testing of the null hypothesis $H_0:\rho = 1$ (see Section \ref{sec:TheNull}).  For clarity, we shall refer to such an exact/point-wise/non-interval null hypothesis as {\it point} null. Restrictions over the domain of $\rho$ are also adopted in \citep{geweke1988secular} and \citep{dejong1991reconsidering}, with the latter also providing an empirical analysis and a comparative study over the classical approach on the Nelson-Plosser data. The arbitrarity in selecting the restricted domain for the autoregressive parameter and the values of the statistics supporting a unit root decision are pointed out and criticized in \citep{sowell1991dejong} and \citep{phillips1991criticize}.

Opposed to the use of non-flat priors like Normal-Wishart conjugates, which are known to be informative about the properties of the model \citep{zellner1971introduction,phillips1991criticize} and that correspond to a prior belief that explosive roots are unlikely when centered around the unit root \citep{uhlig1994macroeconomists}, the choice of flat priors is generally attractive. This is due to that flat priors often appear as a suitable approach for attributing a degree of \enquote{neutrality} or \enquote{objectivity} to Bayesian analyses, while being convenient in terms of computations, often leading to algebraic solutions \citep[e.g.][]{zellner1971introduction}. The early Bayesian inference as of \citep{sims1988bayesian} strongly favors the above approach, but the use of flat priors is not innocuous as it may appear.
\cite{phillips1991criticize} raised concerns on uniform priors since the inference on $\rho$ is conditional on the observed sample moments and sufficient statistics, which depend on the value taken by $\rho$ and are radically different for $\rho= 1$ and $|\rho|<1$. 
The use of flat priors does not correspond to uninformativeness and indeed is shown to downweight large values of $\rho$, i.e. of unit roots and explosive processes. This is because when $|\rho|$ is large, the data is more informative about $\rho$ and treating all the values of $\rho$ as equally-likely implicitly corresponds to downweighting large values of $\rho$.
Consequently, the testing strategy based on regression with or without unit roots taken with the same likelihood irrespectively of the value of $\rho$ is inadequate. Furthermore, \cite{phillips1991criticize} finds that discrepancy in the results between standard and Bayesian methods in unit root testing for macroeconomic US time series is largely due to the use of the misleading flat prior. 

As an objective setting \citep{phillips1991criticize} proposed a Jeffreys' prior \citep{jeffreys1946invariant,perks1947some}. Jeffreys' priors (often called \enquote{ignorance priors}) are defined up to a proportionality factor as $\propto \sqrt{|i|}$, where $|i|$ is the determinant of the expected Fisher information matrix $i$. Jeffreys' priors render the posterior invariant under one-to-one reparametrizations and enjoy a number of desirable properties \citep{ly2016harold}. Such priors are often interpreted as reflecting the properties of the sampling process and emphasizing data evidence, being interpretable as equivalent to the information that a typical single observation, on average, provides \citep{xia2012bayesian}.
Yet, the general use of the Jeffrey's prior with respect to the flat prior and its ability to convey a state of \enquote{ignorance} about the existence of a unit root has been readily argued in \citep{leamer1991comment}. Furthermore, in an extensive Monte Carlo study \cite{kim1991flat} find that the above approach favors big values of $\rho$ distorting the sample evidence. \citep{uhlig1994macroeconomists} finds however that for the univariate AR(1) model the major differences between flat and informative priors are limited to the explosive regions. Further details on the approach of \citep{phillips1991criticize} are provided in Appendix \ref{sec:appendix_phi}.

\subsection{The null hypothesis} \label{sec:TheNull}
As argued in \citep{schotman1991bayesian}, a genuine and exclusive interest in the unit root hypothesis formally corresponds to testing the null $\rho = 1$. Therefore, this hypothesis should be preferred for a Bayesian treatment of unit root inference. If a continuous prior density for $\R$ is to be adopted, the probability associated with $\rho = 1$ is zero and so its posterior probability \citep[e.g.][]{zivot1994bayesian}. In other words, testing a point null $\rho = 1$  -- which corresponds to the only exact hypothesis on the existence of a unit root -- is not trivial. 
Indeed, a continuous prior $\pi\br{\R}$ requires a full Bayesian analysis for retrieving the corresponding posterior $\pi \br{\R|x,x_0} \propto L\br{x|\R,x_0} \pi\br{\R}$, for some data $x$ and initial value $x_0$. 

A posterior confidence interval at the probability level $\alpha$ is a subset $C_\alpha$ such that
\begin{equation*}
    \Prob \br{C_\alpha|x,x_0} = \int_{C_\alpha}\pi\br{\R|x,x_0} \, d\R \geq 1-\alpha \, \text{,}
\end{equation*}
where typically the subset $C_\alpha$ is chosen is such a way that its size is minimal (highest posterior density criterion). 
The unit root hypothesis can be rejected by checking whether $\R=1$ does not belong to the posterior confidence interval $C_\alpha$. 
However, it is not unlikely for the posterior to be bimodal \citep[e.g.][]{kim1991flat} and $C_\alpha$ may result in a disconnected set. Since the interest is around the unit root, an alternative is to redefine $C_\alpha$ so that
\begin{equation*}
    \Prob \br{C_\alpha|x,x_0} = \int_{\R_{\inf}}^{\R_{\sup}} \pi\br{\R|x,x_0} \, d\R \geq 1-\alpha \, \text{,}
\end{equation*}
with $\R_{\inf}$ ($\R_{\sup}$) being  any convenient truncation point of the likelihood or $-\infty$ ($+\infty$). Alternatively, one can consider the probability
\begin{equation} \label{eq:int_1_inf}
    \Pr\br{\R \geq 1|x,x_0} = \int_{1}^{+\infty} \pi \br{\R|x,x_0} \, d\R
\end{equation}
and decide to reject the unit root when its value is below a certain threshold, say 5\%.
This however is not anymore a test on a point null hypothesis as now $H_0$ jointly covers the unit and explosive roots cases:
\begin{equation*}
    H_0: \R \geq 1 \qquad \qquad \qquad H_1:\R < 1 \, \text{.}
\end{equation*}
This is the setting adopted by \citep[e.g.][]{koop1994objective,dejong1991reconsidering,phillips1991criticize}. 

The continuity problem and the impossibility for testing a point null could be technically resolved
by using a discontinuous prior. 
In fact, a natural solution is to give $\rho = 1$ a positive probability $\pi_0$ and assign to the values of $\rho$ over some interval $S$ the density $\br{1-\pi_0}\pi\br{\rho}$, where $\pi\br{\rho}$ is a proper prior on $S$ \citep[as e.g. in][]{schotman1991bayesian,dejong1991reconsidering}.
As in \citep{schotman1991bayesian,zivot1994bayesian} a common choice for $S$  is the stationary region $|\rho| < 1$.
With this procedure, a discontinuous prior can be easily adopted and the following hypotheses can be tested
\begin{equation*}
    H_0: \R = 1 \qquad \qquad \qquad H_1:\R \in S \, \text{.}
\end{equation*}
In general, $H_1$ is not a generic non- unit root alternative since explosive processes are ruled out, and neither a properly stationary alternative, since the lower bound of $S$ could potentially be either within or outside the stationary region. 
Furthermore, is it possible to draw a data-driven criterion for selecting $S$ in the testing procedure (see Appendix \ref{sec:appendix_svd}). In this case, the specific formulation of the alternative depends on the OLS estimate of $\rho$ and on the sample size. It is therefore the test itself that defines the possible values of $\rho$ under $H_1$, and such implicitly imposed restrictions on $S$ are de facto analogous to adopting a strong prior. 

\subsection{Other issues}
In addition to the above difficulties, the whole inferential procedure is remarkably sensitive on model formulation, and the effect of nuisance parameters is not negligible.

\cite{schotman1991bayesian} warn on the effects of extending the AR(1) model with trends and intercepts since the inclusion of such elements changes with (or implicitly corresponds to) prior beliefs. Their study shows that the use of Jeffreys' priors as in \citep{phillips1991criticize} downweights the unit root hypothesis relative to a flat prior in models with trend and intercept. By using the alternative reduced-form parametrization of the AR model with trend \cite{schotman1991bayesian} are able of explaining the observed bias towards stationarity under flat priors. 

Lately, \cite{phillips1991criticize} explains that such behavior is dependent on the initial value and that conditioning the likelihood on the $x_0$ resolves the issue. The apparently secondary role played by the initial value is thus shown to have an enormous impact inference. Analyses in this regard can be found in \citep{zivot1994bayesian} and \citep{lubrano1995testing}. The latter shows that the treatment of the first observation produces results that are more or less in accordance with the classical results and that e.g. the fixed or random treatment of $x_0$ does make a difference when an intercept is included or not (opposed to the simple AR(1) as outlined in \citep{zellner1971introduction}). Following \citep{thornber1967finite}, \cite{lubrano1995testing} extends the discussion suggesting the use of uninformative Beta densities.

Further issues related to the choice of priors are interactions/correlations between the different elements in multivariate parameters (i.e. appropriate priors specifications with non-diagonal covariance matrices), their often improper nature, and computational issues \citep[e.g.][]{zivot1994bayesian}. Also, the conceptual problem of adopting priors that are irrespective and insensible to the sampling frequency of the observations has been pointed out e.g. in \citep{leamer1991comment} and \citep{sims1991comment}.\\

\subsection{Some recent developments}
The literature review would be certainly incomplete without references to more recent works. Though the major advances in the theory of Bayesian unit root testing come from the 20\textsuperscript{th} century, Bayesian unit root testing keeps being an area of active research.

Closely related to the above literature is the Full Bayesian Significance Test (FBST) of \cite{de1999evidence}. This procedure allows to test for the point null unit root hypothesis, has no limiting requirements on the prior, allows for flexible error distributions, applies to small sample sizes, and is invariant with respect to models' parametrization. 
Though attractive, the FBST test has not gained popularity in Bayesian unit root testing as the only application is that of \citep{diniz2011unit} where the FBST performance is compared against the Augmented Dickey-Fuller but not against existing alternative Bayesian methods, in both their simulation study and application.

Recently also unit root testing in Stochastic Volatility (SV) models where the underlying volatility process is unobservable has attracted several research contributions. \cite{so1999bayesian} develop a Monte Carlo Markov Chain approximation of the odds for certain SV models. Their method is improved in \citep{li2010new} with a more robust algorithm and increased test power. Extensions with leverage effects within an SV model on an AR(1) process are considered in \citep{Li_2012}. \cite{Kalayl_o_lu_2009} focus on the role played by priors in unit root Bayesian inference in SV models. They introduce a class of non-informative priors to develop a testing procedure that is feasible based on Gibbs-sampled approximations of the posterior but unpractical for adopting Bayes factors. Simulation methods for approximating posterior credibility intervals are also adopted in \citep{kalayliouglu2013bayesian}, where correlations between returns' series errors and the latent SV with potential unit roots are allowed to be non-zero. Extensions for heteroskedasticity are considered in \citep{chen2013bayesian}. Severe distortions in the size of the Dickey-Fuller test statistics under unit roots in an AR(1) with SV are reported in an extensive simulation study in \citep{zhang2013unit}, where a Bayesian testing approach is introduced as a remedy. Developments over more complex dynamics over the simple AR(1) model include non-normal innovations \citep{Hasegawa_2000}, polynomial trends \citep{Chaturvedi_2005}, nonlinear smooth transitions \citep{Chen_2015} and structural breaks \citep{Park_2016,Vosseler_2016}, also in panel data \citep{Kumar_2018}.

Not strictly relevant for our analysis, yet closely related to Bayesian unit root literature, are the dozens of research works on Bayesian cointegration testing, appeared since \citep{koop1991cointegration}.

\section{Evidence} \label{sec:evidence}
Consider a random variable $X$ with density parametrized over $\theta \in \Theta$. The hypothesis testing problem we are concerned about consists of deciding among the null hypothesis $H_0:\T = \T_0$ and the alternative $H_1: \T \neq \T_0$. This is achieved by considering suitable measures of evidence of a hypothesis against the other, such as the widespread p-value, or Bayes factors and Bayesian posterior probabilities.

\subsection{P-value}
Let us denote by $T\br{\cdot}$ a test statistics, and by $t = T\br{x}$ its value when data $X=x$ is observed. The null hypothesis $H_0$ is rejected in favor of the alternative $H_1$ if $T\br{x}$ is more extreme than one would expect if $H_0$ was true. By choosing a significance level $\alpha$, $H_0$ is rejected when the probability of $T\br{X}$ being greater than $T\br{x}$ is small (i.e. lower or equal to $\alpha$), given that $H_0$ is true. Formally, the hypothesis $H_1$ is accepted if
$$
\Pr\br{\abs{T\br{X}}\geq T\br{x} \lvert H_0} \leq \alpha \, \text{,}
$$
that is, extreme values of the statistics are deemed to provide evidence against $H_0$. 

While it is straightforward to identify in higher values of $|t|$ stronger evidence against the null hypothesis, the problem of evaluating the strength of evidence for $H_0$ against $H_1$ is left open. Frequentists use a scale of evidence set from Fisher's work in the 1920s: a common interpretation is that $\alpha=0.99$ corresponds to very strong evidence, $\alpha=0.95$ to strong evidence and so on, with neutral evidence at around $\alpha=0.9$. In other words, the higher the evidence the lower the Type I error of rejecting the true hypothesis.
Bayesian literature provides different answers to the problem, based on quantities known as {\it Bayes factor} and {\it posterior odds}.

\subsection{Posterior odds for a point null}
Considering the unknown model probabilities as random, the Bayes' rule yields posterior probabilities given the observed data. For the hypothesis $H_1$ (and analogously for $H_0$):
$$
\Prob\br{H_1|x}= \frac{\Pr\br{x|H_1}\Prob\br{H_1}}{\Prob\br{x}} \, \text{,}
$$
where $\Prob\br{H_1}$ is the prior probability of hypothesis $H_1$ being true, and $\Prob\br{x} = \Pr\br{x|H_0}\Prob\br{H_0} + \Pr\br{x|H_1}\Prob\br{H_1}$ is the marginal density of $X=x$.
The quantity $\Pr\br{x|H_1}$ is referred to as {\it marginal likelihood} or {\it marginal probability} (of the data) under $H_1$.
In this context, the language and notation encountered in the literature can be heterogeneous and slightly abused as it is common to indistinctly
use the terms (joint) ``density", ``likelihood", (joint) ``probability"
and their corresponding notations, e.g. $f$, $L$, $\Pr$. Here we adopt the notation and terminology of \citep{kass1995bayes}.

By the law of the total probability the marginal probability $\Pr\br{x|H_1}$ is obtained by marginalizing out the parameter $\T_1$ under $H_1$:
\begin{equation} \label{eq:mgx}
\Pr\br{x|H_1} = \int \Pr\br{x|\theta_1,H_1} \pi\br{\theta_1|H_1} d\theta_1 \, \text{,}
\end{equation}
where $\pi\br{\T_1|H_1}$ is a continuous density.
From a frequentist perspective, $\pi\br{\T_1|H_1}$ is a mere weight function to allow the computation of the average likelihood. For a Bayesian, $\pi\br{\T_1|H_1}$ would be the {\it prior} density for $\T_1$ conditional on $H_1$ being true \citep{berger1987testing}. Note the difference between the prior {\it probability} of a hypothesis or model being true opposed to the prior {\it density} referring to its corresponding parameter.

The ratio of the posterior probabilities for the two hypotheses is referred to as  {\it posterior odds ratio}, or {\it posterior odds}:
$$
\frac{\Prob\br{H_0|x}}{\Prob\br{H_1|x}} = \frac{\Pr\br{x|H_0}}{\Pr\br{x|H_1}} \, \frac{\Prob\br{H_0}}{\Prob\br{H_1}} \, \text{.}
$$
Analogously, one may define the {\it prior odds} as $\Prob\br{H_0}/\Prob\br{H_1}$.
Posterior odds quantify the evidence of $H_0$ over $H_1$ after data $x$ has been observed.  
On the other hand, prior odds do not convey any evidence as they solely quantify the prior plausibility of $H_0$ over $H_1$ before any data is observed. 
The interpretation of odds ratios is straightforward as they correspond to simple probability ratios. Odds ratios $K_1$ greater than one (or $\log K_1>0$) stand for evidence in favour of the null hypothesis, with a corresponding probability $K_1/\br{1+K_1}$. 
This is aligned with the general Bayesian rationale. 
After observing $x$, the prior probability $\Prob\br{H_0}=\pi_0$ and corresponding prior odds $K_0 = \pi_0/\br{1-\pi_0}$  reflecting the prior belief on $H_0$ being the true model 
are updated into the posterior odds $K_1$ and the corresponding new posterior probability for $H_0$ being true.

For a point hypothesis $H_0: \theta = \theta_0$, the assignment of a positive probability will be rarely thought possible for $\theta = \theta_0$ to hold exactly: this is to be understood as a realistic approximation of the hypothesis $H_0: |\theta-\theta_0|\leq b$, for some small $b$, so that $\pi\br{\T_0|H_0}$ in fact represents the prior probability assigned to $\lbrace |\theta-\theta_0|\leq b \rbrace$ \citep[see][]{berger1987irreconcilability}. The way to depict such prior is through a smooth density with a sharp peak around $\theta_0$.

\subsection{Bayes factors}
The focus of this paper is on the ratio 
$$
B_{01} = \frac{\Pr\br{x|H_0}}{\Pr\br{x|H_1}} \, \text{,}
$$
referred to as {\it Bayes factor}. With this definition, the posterior odds for a null hypothesis $H_0$ and the alternative $H_1$, as discussed so far, can be written as:
$$
\frac{\Prob\br{H_0|x}}{\Prob\br{H_1|x}}  =  B_{01} \frac{\Pr\br{H_0}}{\Pr\br{H_1}} \text{.}
$$
Bayes factors can be interpreted as either the ratio quantifying the plausibility of observing the data $x$ under $H_0$ over $H_1$, or as the degree by which the observed data updates the prior odds $\Prob\br{H_0}/\Prob\br{H_1}$.
With respect to the posterior odds which involve prior probabilities on the hypotheses, the interest in Bayes factors arises from the fact that they appear as actual odds implied by the observed data only. Moreover, Bayes factors are of attractive interpretation since they can be viewed as likelihood ratios obtained by averaging likelihoods $\Pr\br{x|\T_k,H_k}$ across $\T_k|H_k$, with weights $\pi\br{\T_k|H_k}$, $k=\{0,1\}.$

Not less importantly, the calculation of the Bayes factor requires the prescription only of the prior distributions $\pi\br{\T_k|H_k}$, while the full Bayesian analysis leading to posterior odds requires the additional specification of the prior probabilities $\Prob\br{H_k}$, for $k=\{0,1\}$.
The interpretation from above still applies: a Bayes factor of e.g. $1/10$ means that $H_1$ is supported by an evidence 10 times as high as $H_0$ is.
Furthermore, under the appealing ``neutral" choice $\Prob\br{H_0} = \Prob\br{H_1} = 1/2$, Bayes factors coincide with posterior odds, further enforcing $B_{01}$ as a suitable alternative to p-values.
Similar to the frequentist approach where decisions are based on the critical level $\alpha$, the higher the Bayes factor the stronger the evidence in favor of $H_0$ over $H_1$. \citep{jeffreys1961theory} provides a scale for interpreting $B_{01}$ as the degree to which $H_0$ is supported by the data over $H_1$, with ratios greater than 100 being decisive. A comparison between the Bayesian and frequentist decision scales can be found in \citep{efron2001scales}.

\subsubsection{Bayes factors for unit root testing}
We show how the above description on Bayes factor and posterior odds practically applies to test the set of hypotheses
$$
    H_0: \R = 1 \qquad \qquad \qquad H_1: \R \in S \text{.}
$$
Here we extend the discussion to any generic AR-like model, holding the usual interpretation of $\R$ as the autoregressive parameter on interest.
First, to compare the average likelihood of a model over the complementary region through posterior odds as outlined above, we assign a certain positive weight $\pi_0$ to the point null hypothesis $H_0$ and share the complement $\br{1-\pi_0}$ over the interval $S$ relevant under $H_1$.
Second, as a general case the likelihood function is parametrized over a rich $K$-dimensional parameter $\{ \R, \T \}$ defined over some appropriate set $\{ S,\Theta \}  \in \mathbb{R}^K$. Therefore, the computation of the Bayes factor in this setting involves a multidimensional marginalization over the elements of $\theta$, thus posteriors odds read:
\begin{equation} \label{eq:K1_long}
K_1 = \frac{\pi_0}{1-\pi_0} \frac{ \int_\Theta L\br{x|\R = 1,\theta, x_0} \pi\br{\theta|\rho=1} d\theta}{ \int_S \int_\Theta L\br{x|\R,\theta,x_0}\pi\br{\theta|\rho}\pi\br{\R} \, d\theta d\R} = \frac{\Pr\br{\R =1| x}}{ \Pr\br{\R \in S| x}}\, \text{.}
\end{equation} 
Eq.\eqref{eq:K1_long} embeds a feasible and largely adopted specification for the prior over $\br{\R,\T}$  that imposes conditional independence on the parameters, allowing for a convenient factorization of the joint prior as a product of conditionally independent factors, that is $\pi\br{\R,\T} = \pi\br{\T|\R} \pi\br{\R}$. Rather than the above probability notation ($\Pr$) typical in introductory discussions on Bayes factors, here we use the likelihood notation ($L$) as it is more common in the related econometric literature.

For inference on the simple AR(1) process
$$
    x_t = \rho x_{t-1} + u_t \, \text{,}
$$
$K=2$, as $\T$ generally includes the unknown variance $\S$ of the innovations. In this case, posterior odds have the simple form
\begin{equation*}
    K_1 = \frac{\pi_0}{1-\pi_0} \frac{ \int_{0}^{\infty} L\br{x|\R =1, \S, x_0}\pi\br{\S} d\S}{\int_S \int_{0}^{\infty} L\br{x|\R,\S, x_0} \pi\br{\S} \pi\br{\R} \, d\S d\R}  \, \text{.}
\end{equation*}
The conditional independence assumption among the parameters is generally not very restrictive and commonly extended to unconditional independence, thus $\S$ is provided with its own prior $\pi\br{\S|\rho} = \pi\br{\S}$.
Further information can be found e.g. in \citep[][among the others]{schotman1991bayesian, zivot1994bayesian,zellner1980posterior}. 

In the following section we show how to approximate the general Bayes factor involved in Eq.\eqref{eq:K1_long} with a friendly form of moderate error that neither requires integration nor the priors to be specified.

\section{Hypothesis testing with BIC}  \label{sec:approximations}
\subsection{Laplace approximation}
For $k = \{0,1\}$ consider the densities $\Pr\br{x|H_k} = \int \Pr \br{x|\T_k,H_k} \pi \br{\T_k|H_k} d \T_k$ involved in the definition of Bayes factors.
Be $\T_k$ the parameter under $H_k$, $\pi\br{\T_k|H_k}$ its prior, and $\Pr\br{x|\T_k,H_k}$ the prior density of $x$ given the values of $\T_k$. $\T_k$ in general represents a vector parameter with dimension $d_k$. 
In the following, we shall refer to the marginal probability of the data (or marginal likelihood) $\Pr\br{x|H_k}$ as $I$ and adopt a simplified notation where we drop $k$ and rewrite the marginal likelihood as
$$
I = \int \Pr(x|\T,H) \pi \br{\T|H} d \T \, \text{.}
$$
Except for some elementary cases where the above integral can be evaluated analytically, the computation of the marginal likelihood is intractable and requires numerical methods. 
In fact, analytic solutions for $I$ are limited to exponential family distributions and conjugate priors, including normal linear models \citep[e.g.][]{degroot2005optimal,zellner1971introduction}. A general description of the different approaches for evaluating $I$ with numerical methods is provided in \citep{evans1995methods}.

To recover a first useful approximation for $I$, assume that the posterior density, proportional to $\Pr(x|\T,H) \pi\br{\T|H}$, is peaked around its maximum $\Tilde{\theta}$, which is the posterior mode. This is generally the case for large samples if the likelihood function of the data $\Pr(x|\T,H)$ is peaked around its maximum $\hat{\theta}$ \citep{kass1995bayes}.
Let $g\br{\T} = \log \br{\Pr(x|\T,H)\pi\br{\T|H}}$ and consider its Taylor expansion around $\tilde{\T}$: $g(\T) = g(\TT)+ (\T-\TT)^Tg'(\TT)+\sfrac{1}{2} (\T-\TT)^Tg''(\TT) (\T-\TT) + o(||\T-\TT||^2)$. Since $g'(\TT) = 0$ as $g$ reaches its maximum at $\TT$, it follows that:
\begin{align}\label{eq:ILongIntegral}
 I &= \int \text{exp}[g(\T) ] d\T \,\, \approx \text{exp}[g(\TT) ] \int \text{exp} \left[ \sfrac{1}{2} (\T-\TT)^Tg''(\TT) (\T-\TT) \right] d\T \, \text{,}
\end{align}
where we recognize in the integrand the kernel of a generic $d$-dimensional multivariate normal distribution with mean $\TT$ and {\it covariance} matrix $\tilde{\Sigma}$. $\tilde{\Sigma}$ also corresponds to minus the inverse Hessian matrix of the second order derivatives of $g(\T)$ evaluated at $\T = \tilde{\T}$, i.e. $\tilde{\Sigma}^{-1} = -g''(\TT)$. The integrand in Eq.\eqref{eq:ILongIntegral} therefore equals  $\br{2\pi}^{d/2}\vert \tilde{\Sigma} \vert ^{1/2}$, from which the following approximation is known as {\it Laplace approximation} \citep[e.g.][]{konishi2008information}:
\begin{equation}\label{eq:ILaplace}
    \tilde{I} = \br{2\pi}^{d/2} \vert \Tilde{\Sigma} \vert ^{1/2} \Pr(x|\TT,H) \pi (\TT|H)
    \, \text{.}
\end{equation}
In particular, as $n$ diverges $I=\tilde{I}(1+O\br{n^{-1}})$, see e.g. \citep{tierney1989fully,kass1991laplace}. 
Eq.\eqref{eq:ILaplace} can be applied to any regular statistical model and stands as a viable general approach for evaluating the marginal likelihoods involved in the definitions of Bayes factors with an approximation error of order $O\br{n^{-1}}$.
\cite{slate1994parameterizations} discusses requirements on the sample size for reaching posterior normality, and the accuracy of Laplace's method has been more generally investigated in \citep[e.g.][]{Bradley1978assessing,kass1992approximate}. An empirical rule is provided in \citep{kass1995bayes}: sample sizes of at least $5d$ provide a satisfactory accuracy in well-behaved problems, with $20d$ applicable in most situations.

The use of Eq.\eqref{eq:ILaplace} is impractical since $\tilde{\T}$ refers to the posterior mode and $\tilde{\Sigma}$ to the negative inverse Hessian of $g\br{\T}$, while maximum likelihood estimates and information matrices are of common use and generally readily available as standard outputs in any statistical software.
Indeed, a variation over Eq.\eqref{eq:ILaplace} that has attracted much attention uses the maximum likelihood estimator $\hat{\T}$, applies to large samples where $\tilde{\T} \approx \hat{\T}$, and relies on the covariance matrix $\hat{\Sigma}$ so that $\hat{\Sigma}^{-1}$ corresponds to the {\it observed } information matrix, i.e. the negative Hessian of the log-likelihood evaluated at $\hat{\theta}$ \citep{tierney1989fully,kass1992approximate}:
\begin{equation}\label{eq:IMLE}
     \hat{I} = \br{2\pi}^{d/2} \vert \hat{\Sigma} \vert ^{1/2} \Pr(x|\hat{\T},H) \pi(\hat{\T}|H)
     \, \text{.}
\end{equation}
The relative error in this case is still the best rate $O\br{n^{-1}}$. If one replaces the observed information matrix with the {\it expected} information matrix $i$, the asymptotic error rate moves to the larger order $O\br{n^{-1/2}}$. The expected information matrix $i$ is a $d \times d$ matrix whose $\br{h,k}$ element is
$$
-\mathbb{E}\left[ \left. \frac{\partial \log \Pr\br{x_1|\T,H}}{\partial \T_h  \partial \T_k}  \right| _{\T = \hat{\T}} \right] \, \text{,}
$$
and the expectation is taken over $x_i$ with $\T$ held constant. 
Therefore, in large samples the observed information matrix $\hat{\Sigma}^{-1}$ can be approximated based on the expected information matrix $\hat{\Sigma}^{-1}  \approx ni$, and $n^d  i =\vert \hat{\Sigma}^{-1} \vert = \vert \hat{\Sigma} \vert ^{-1} $. With this substitution Eq.\eqref{eq:IMLE} rewrites as:
$$
\hat{I} = \br{2\pi}^{d/2}n^ {-d/2}\vert i \vert ^{-1/2} \Pr(x|\hat{\T},H) \pi(\hat{\T}|H) \, \text{,}
$$
from which
\begin{align} \label{eq:logI_O0.5}
 \log I = \log \Pr(x|\hat{\T},H) + \log \pi(\hat{\T}|H) + \frac{d}{2} \log\br{2\pi} - \frac{d}{2}  \log n - \frac{1}{2} \log |i| + O\br{n^{-1/2}} \, \text{.}
\end{align}
Note that the prior density  $\pi\br{\T|H}$ needs to be fully specified, as it is involved throughout the approximation procedure. 

This leads to the conclusive approximation form that does {\it not} involve prior densities:
\begin{align} \label{eq:log_O1}
  \log I = \log \Pr(x|\hat{\T},H) - \frac{d}{2} \log n+O\br{1} \, \text{.}
\end{align}
This last approximation is in virtue of the fact that in Eq.\eqref{eq:logI_O0.5} besides 
$\Pr(x|\hat{\T},H)$ and  $\log n$ which are respectively of order $O\br{n}$ and $O\br{\log n}$, all the remaining terms are of order $O\br{1}$ or lower. From Eq.\eqref{eq:log_O1} we have that the marginal likelihood is thus equal to the maximized likelihood $\Pr(x|\hat{\T},H)$ minus a correction term where the approximation is $O\br{1}$.
Even though the $O\br{1}$ term does not vanish, because all the other terms tend to infinity as $n$ increases, the error is dominated and vanishing as a proportion of $I$. \cite{raftery1995bayesian} shows that in reality the error term is not as high as one might think, although an $O\br{1}$ error suggests that the approximation is in general quite crude. In fact, the error can be of a smaller order of magnitudes given a reasonable choice of the prior.

As a remark, the definition of the sample size $n$ should reflect the rate at which the Hessian matrix of the log-likelihood grows, i.e. satisfactory for the approximation $\hat{\Sigma}^{-1} \approx ni $. This $n$ turns to be the number contributions to the summation appearing in the definition of the Hessian \citep{raftery1995bayesian,kass1995bayes} -- e.g. in survival analysis, $n$ would match the number of non-censored observations rather than the total number of observations.

\subsection{BIC Approximation of the Bayes factor}

The above discussion provides the basis for the following approximation of the Bayes factor. Hereafter we, focus on the case where the null hypothesis is nested. That is, we assume some parametrization under $H_1$ of the form $\T_1 = \br{\R,\beta}$ such that $H_0$ is obtained from $H_1$ by imposing the restriction $\R = \R_0$ for some $\R_0$. 
Both $\R$ and $\beta$ can be vectors. 
Let $\T_1$ denote the parameter under $H_1$  with prior $\pi\br{\T_1|H_1} = \pi\br{\R,\beta|H_1}$, and for $H_0: \R=\R_0$ let its prior be $\pi\br{\T_0|H_0} = \pi\br{\beta|H_0}$.

Based on Eq.\eqref{eq:IMLE}, by applying the definition of the Bayes factor to the log-ratio of the marginal likelihoods, one obtains:
$$
2\log B_{10}  \approx \Lambda + \log |\hat{\Sigma}_1| -\log|\hat{\Sigma}_0|  +2\log \pi(\hat{\T}_1\vert H_1)  -2\log \pi(\hat{\T}_0 \vert H_0)  +\br{d_1-d_0}\log\br{2\pi}
$$
where  $\Lambda_{10} =  2(\log \Pr(x|\hat{\T}_1,H_1)- \log \Pr(x|\hat{\T}_0,H_0))$ 
corresponds to the log-likelihood ratio statistics with $d_1-d_0$ degrees of freedom. 
Refer to \citep{raftery1996approximate} for an additional discussion, and for the approximation of the Bayes factor under to Eq.\eqref{eq:ILaplace}.
On the other hand, based on the approximation in Eq.\eqref{eq:log_O1}, one obtains:
\begin{align} \label{eq:approx3}
    2 \log B_{10} &\approx \Lambda_{10} - \br{d_1-d_0} \log\br{n} =2S_{10} \, \text{,}
\end{align}
where
\begin{align*}
2S_{10} &= 2\log \Pr (x|\hat{\T}_1, H_1) - 2\log \Pr (x|\hat{\T}_0, H_0)-\br{d_1-d_0} \log\br{n}\\
        & = \left[ d_0\log\br{n} - 2\log \Pr (x|\hat{\T}_0, H_0)\right] - \left[ d_1\log\br{n} - 2\log \Pr (x|\hat{\T}_1, H_1) \right]
\, \text{.}
\end{align*}
The following consistency result for $n \to \infty$ known as Schwarz criterion 
\begin{equation} \label{eq:SchwartzCrit}
\br{S_{10}-\log B_{10}}/\log B_{10} \to 0 \, \text{,}
\end{equation}
is attractive as it establishes $S_{10}$ as a standardized quantity to be used even when the priors are hard to set and a useful reference quantity in scientific reporting \citep{kass1995bayes}. The $O\br{1}$ error implies that even in large samples $S_{10}$ does not lead to the correct value in absolute terms, the error does go to zero in terms of proportion with respect to the actual log of the Bayes factor \citep[e.g.][]{kass1995bayes,raftery1995bayesian,raftery1996approximate}.
Importantly, for certain classes for priors the error of approximation reduces to $O\br{n^{-1/2}}$. One class is that of Jeffrey's priors with a specific choice of the constant preceding them, another class is that of unit information priors \citep{raftery1995bayesian,wasserman2000bayesian,wagenmakers2007practical}. With respect to subjectively determined priors, a surprisingly good agreement between the Schwarz criterion and actual Bayes factors is observed in \cite{kass1995reference}.
In general when the sample size $n$ is sufficiently large, the approximation is very satisfactory for most of purposes and is of widespread use, including applications in psychology \citep[e.g.][]{wasserman2000bayesian}, ecology \citep[e.g.][]{aho2014model}, and computer vision \citep[e.g.][]{stanford2002approximate}. 
\cite{kass1995reference} further show that for the intuitive and reasonable choice of the unit information prior $\text{exp}\br{S_{10}}/B_{10} \to 1$ with an error of order $O\br{n^{-1/2}}$. This provides a direct interpretation of the Schwartz criterion in terms of Bayes factors and evidence.

For a given model $k$, recall the definition of the Bayesian Information Criterion (BIC):
\begin{equation} \label{eq:BICdef}
 \text{BIC}_k = d_k \log n - 2 \log \Pr(x \vert \hat{\T}_k,H_k) \, \text{.}
\end{equation}
It is easy to recognize that the right side of Eq.\eqref{eq:log_O1} is closely related to BIC as $- 2 \log I = \text{BIC}+O\br{1}$, and that $2S_{10} = \text{BIC}_0 - \text{BIC}_1 = \Delta \text{BIC}_{01} $. Eq.\eqref{eq:approx3} then is equivalent to
\begin{equation} \label{eq:log_BIC}
\log B_{10} \approx  \frac{1}{2}\Delta \text{BIC}_{01} \, \text{.}
\end{equation}
Eq.\eqref{eq:log_BIC} establishes $\Delta \text{BIC}_{01}$ as an approximate measure of the log-evidence in support of the hypothesis $H_1$ over $H_0$. 
We shall refer to either Eq.\eqref{eq:log_BIC} or its exp-version $B_{10} \approx \exp \br{\sfrac{1}{2} \text{BIC}_{01}}$ as the BIC approximation of the Bayes factor.

For the above BIC approximation the approximation error is generally $O\br{1}$, but in virtue of the Schwartz criterion $\br{\Delta \text{BIC}_{10}-\log B_{10}}/\log B_{10} \to 0 $ the error approaches zero as a proportion of the Bayes factor. As discussed above, it can further reduce to $O\br{n^{-1/2}}$ for certain priors.
Eq.\eqref{eq:log_BIC} formally justifies the extensive practice of model selection based on the smallest BIC value. Indeed, the higher the evidence in support of model 1, i.e. the higher the log-Bayes factor, the more positive $\Delta \text{BIC}_{01}$ and the smaller $\text{BIC}_1$ with respect to $\text{BIC}_0$.

In the context of linear models with normal errors, BIC rewrites in the alternative convenient form
\begin{equation*}
    \text{BIC}_k = n\log\br{1-R^2_k}+d_k\log n \, \text{,}
\end{equation*}
with $R^2$ being the usual R-squared. The proportion  $1-R^2_k$ of the variance that model $k$ fails to explain relates to the sum of squares table through the equivalence $1-R^2_k = SSE_k-SS_{\text{total}}$, where $SSE_k$ is the sum of squared errors for model $k$ and $SS_{\text{total}}$ the total sum of squared errors. This leads to the following expression:
\begin{equation} \label{eq:BIC_SSE}
    \Delta \text{BIC}_{01} = n \log{\frac{SSE_0}{SSE_1}}+\br{d_0-d_1}\log\br{n} \,\text{.}
\end{equation}
Applied examples on the use of Eq.\eqref{eq:BIC_SSE} are provided e.g. in \citep{wagenmakers2007practical,masson2011tutorial}.

Furthermore, for nested models such that $d_1-d_0 = 1$, $\Lambda_{10} \approx t^2$ with $t$ being the t-statistics for testing the significance of the parameter in model $1$ that is set to zero in model $0$, and $\Lambda_{10}$ the corresponding likelihood ratio statistics. From Eq.\eqref{eq:log_BIC}:
\begin{equation}\label{eq:BIC_t_approx}
    2\log B_{10} \approx \Delta \text{BIC}_{01} = \Lambda_{10} -\log n  \approx t^2 -\log n \, \text{.}
\end{equation}
This underlines a proportionality between $t$, $\Delta \text{BIC}$ and $B$, which means that the $t$ statistics can be directly translated into BIC and into grades of evidence through Bayes factors \citep{Valen2005bayes}. 
High values of $t$ support the statistical significance of the additional parameter in the full model. In turn, $\text{BIC}_1$ is smaller than $\text{BIC}_0$, so $\Delta \text{BIC}_{01}>0$ and $\log B_{10} >0$, which is indicative of evidence against the reduced model.
By reparametrization, the above extends to hypotheses where an element $\T'$ of  $\theta$ is set to a fixed value (rather than zero), e.g. $H_0: \T' = 1$ is analogous to $H_0:\T'' = 0$ by taking $\T'' = \T'-1$.

\subsection{Unit root testing based on BIC approximation}
The major contribution of this paper is to test for unit roots in financial time series by the BIC approximation of the Bayes factor, Eq.\eqref{eq:log_BIC}. We shall list the major points in favor of this approach.
\begin{itemize}
\item[i.] As reviewed above, the choice of the prior is the principal problem in Bayesian testing of unit roots. On the contrary, our proposed BIC approximation does not require a full specification of the priors. Neither that of the autoregressive parameter, nor of any other parameter. The independence of the BIC approximation on prior specifications is also attractive from the point of objectivity in Bayesian analysis.

\item[ii.] Bayes factors and posterior odds allow test point nulls on the autoregressive parameter, which can be problematic as shown earlier. This setting is however natural for the BIC approximation. 

\item[iii.] The BIC approximation to Bayes factor is a general procedure that does not depend on the model form or parametrization. Regardless of whether the AR model under investigation has an intercept, a trend component, exogenous regressors, or any richer structure, the BIC procedure applies. In general, the error is $O\br{1}$ but reduces to zero as a proportion of the Bayes factor (cf. Eq.\eqref{eq:SchwartzCrit}).

\item[iv.] Testing based on BIC approximation does not require any integration and does not present major computational issues: it only requires the maximum likelihood estimates, as of definition in Eq.\eqref{eq:BICdef}. 

\item[v.] The applicability of the method depends on the feasibility of the approximation in Eq.\eqref{eq:IMLE}, i.e. on the feasible hypotheses that the data $x$ consist of i.i.d. observations, that the posterior is peaked around its maximum, and that the sample size is sufficient for $\tilde{\T} \approx \hat{\T}$ and $\hat{\Sigma}^{-1}\approx ni$ to be satisfactory approximations. A sample size of about 20 times the number of parameters appears to be generally fair for well-behaved problems where the likelihood is not grossly non-normal. For the simple AR(1) model with unknown innovations' variance, this corresponds to about 40 points, e.g. about two months of daily market data. 
\end{itemize}

Lastly, consider that even if the applicability of the above procedure is quite broad, time series models involved in classical unit root testing applications are broadly of linear form, so the simplified alternative form in Eq.\eqref{eq:BIC_SSE} commonly applies. Also, the point form of the null hypothesis naturally suggests a nested hypotheses structure where the only restriction is on the autoregressive parameter. That is, applications will generally encounter linear model specifications where $d_1-d_0=1$.

\section{Simulation Study}  \label{sec:simulation}
To validate our proposed testing methodology and compare it with some existing alternatives, we develop a Monte Carlo simulation study.
In particular, we simulate 20,000 simple AR(1) processes $x_t = \rho x_{t-1}+u_t$, with $t=\{1, \dots, T\}$ and independent standard normal innovations, by considering different sample lengths  $T =\{ 50,100,200,500,1000,5000 \}$ and different values of the autoregressive parameter $\R = \{ 0.2,0.5,0.8,0.9,0.99,0.999,1 \}$. We test the point null unit root hypothesis $H_0: \R = 1$ 
and summarize in Table \ref{tab:sim_p} the corresponding results. All the tables related to the Monte Carlo study report averages across the simulated samples. 

Table \ref{tab:sim_p} includes probabilities associated with different alternative testing methods. For the approach of \cite{schotman1991bayesian} we both adopt a constant lower integration bound $a$ for the autoregressive parameter fixed to $-1$ (SVD) and the data-driven one (SVD*), see Appendix \ref{sec:appendix_svd}.
As a reference measure, Table \ref{tab:sim_p} includes the average p-value for the Dickey-Fuller test (DF) and the posterior non-stationary probabilities $\Pr_{\R\geq1} = \Prob\br{\R \geq 1|x}$ from \citep{phillips1991criticize}, computed through Eq.\eqref{eq:int_1_inf}. These posterior probabilities are reported only for $T\leq 200$ as larger $T$ quickly drive Eq.\eqref{eq:posterior_philips} below machine precision and integration turns problematic: recovering such probabilities for large $T$ is here beyond our scope. For the SVD, SVD* and BIC entries in Table \ref{tab:sim_p}, we report as our main result the probabilities corresponding to the Bayes factors in Table \ref{tab:sim_bf}, since of easier interpretation. Furthermore, assuming prior odds equal to one, these probabilities also interpret as posterior odds.

Our simulation results show a smooth and coherent behavior of the proposed BIC approximation leading to posterior probabilities behaving in accordance with how one might expect from this controlled setting. (i) The acceptance probabilities for the null are progressively higher as $\R$ approaches the unity. (ii) For a fixed $\R$, larger samples reduce the posterior unit root probability for small values of $\R$ while increase it for $\R$ around 1. Indeed larger samples embed higher evidence in support of the true hypothesis, for which we observe increasing posterior probabilities (and log- Bayes factors).
With $\R = 0.8$ and $T = 50$, for the BIC approximation we compute an average posterior probability for $H_0$ of .240, however as $T$ increases the evidence towards the true hypothesis $\R = 0.8$ increases as well, and the posterior probability of the null moves rapidly towards zero (e.g. from $T=200$ ahead). Similarly, for $\R=1$ and $T=50$, the small size of the sample advocates for a stationary dynamics with a considerable 0.211 probability, while at larger $T$ the probability associated with far-from-unity values of $\R$ sharply decreases to zero.

Furthermore, our simulation study leads to posterior probabilities that are also well-aligned across the BIC, SVD and SVD* testing approaches, following the same trends across different values of $\R$ for fixed $T$. The BIC approximation is however leading to probabilities that are not uniformly greater or smaller than those from SVD*. In fact, for small to moderate sample sizes BIC returns higher posteriors probabilities for the unit root hypothesis, while smaller for large $T$ with respect to SVD*. This behavior could be partially explained by the different specifications of the alternative hypothesis. While for SVD the reported probabilities are those of a unit root against a stationary alternative, for BIC the alternative is a generic $H_1:\R \neq 1$. It is thus reasonable that the rejection probabilities are larger for BIC than SVD, as by construction for BIC the feasible parameter space under the alternative $H_1$ is broader. 
Also with respect to posterior probabilities of explosive roots/non-stationarity dynamics $\Pr_{\R\geq1}$ under the ignorance prior, the BIC approximation appears coherent and well-aligned. Though BIC and $\Pr_{\R\geq1}$ probabilities refer to different hypotheses, we observe that indeed the higher the posterior probability of a unit root, the higher the posterior probability supporting the non-stationary option.

As expected, p-values associated with the most classical frequentist Dickey-Fuller test cannot exclude the non-stationary hypothesis with increasing confidence as $\R$ moves towards one. Accordingly, the evidence in support of a unit root is reflected in increasing posterior BIC probabilities. All the above discussion applies as well to the Bayes factors reported in Appendix \ref{sec:appendix_lbf}, where negative signs stand for evidence against the unit root null.

This study confirms an overall very satisfactory behavior of the proposed method with respect to some Bayesian and non-Bayesian alternatives for unit root hypothesis testing. Despite the general $O(1)$ error associated with BIC approximation and its complete independence on the prior specification, our simulated posterior probabilities are well-behaving (i.e. showing desirable smooth monotonicity over $\R$ for $T$ fixed, and the other way around), coherent with their expected behavior, and aligned with the decisions over the null the other approaches suggest.

\begin{table}[ht]
    \centering
    \scalebox{0.9}{
\begin{tabular}{rrrrrrrlrrrr}
      &       &       &       &       &       &       &       &       &       &       &  \\
\multicolumn{6}{c}{\textit{$T=50$}}           &       & \multicolumn{5}{c}{\textit{$T=500$}} \\
\multicolumn{1}{c}{\textit{$\rho$}} & \multicolumn{1}{c}{SVD} & \multicolumn{1}{c}{SVD*} & \multicolumn{1}{c}{BIC} & \multicolumn{1}{c}{DF} & \multicolumn{1}{c}{$\Pr_{\R\geq1}$} &       & \multicolumn{1}{c}{\textit{$\rho$}} & \multicolumn{1}{c}{SVD} & \multicolumn{1}{c}{SVD*} & \multicolumn{1}{c}{BIC} & \multicolumn{1}{c}{DF} \\
\cmidrule{1-6}\cmidrule{8-12}\multicolumn{1}{c}{0.200} & \multicolumn{1}{c}{.000} & \multicolumn{1}{c}{.000} & \multicolumn{1}{c}{.000} & \multicolumn{1}{c}{.001} & \multicolumn{1}{c}{.015} & \multicolumn{1}{c}{} & \multicolumn{1}{c}{0.200} & \multicolumn{1}{c}{.000} & \multicolumn{1}{c}{.000} & \multicolumn{1}{c}{.000} & \multicolumn{1}{c}{.001} \\
\multicolumn{1}{c}{0.500} & \multicolumn{1}{c}{.003} & \multicolumn{1}{c}{.002} & \multicolumn{1}{c}{.004} & \multicolumn{1}{c}{.001} & \multicolumn{1}{c}{.107} & \multicolumn{1}{c}{} & \multicolumn{1}{c}{0.500} & \multicolumn{1}{c}{.000} & \multicolumn{1}{c}{.000} & \multicolumn{1}{c}{.000} & \multicolumn{1}{c}{.001} \\
\multicolumn{1}{c}{0.800} & \multicolumn{1}{c}{.292} & \multicolumn{1}{c}{.124} & \multicolumn{1}{c}{.240} & \multicolumn{1}{c}{.032} & \multicolumn{1}{c}{.244} & \multicolumn{1}{c}{} & \multicolumn{1}{c}{0.800} & \multicolumn{1}{c}{.000} & \multicolumn{1}{c}{.000} & \multicolumn{1}{c}{.000} & \multicolumn{1}{c}{.001} \\
\multicolumn{1}{c}{0.900} & \multicolumn{1}{c}{.686} & \multicolumn{1}{c}{.341} & \multicolumn{1}{c}{.545} & \multicolumn{1}{c}{.111} & \multicolumn{1}{c}{.306} & \multicolumn{1}{c}{} & \multicolumn{1}{c}{0.900} & \multicolumn{1}{c}{.000} & \multicolumn{1}{c}{.000} & \multicolumn{1}{c}{.000} & \multicolumn{1}{c}{.001} \\
\multicolumn{1}{c}{0.990} & \multicolumn{1}{c}{.955} & \multicolumn{1}{c}{.663} & \multicolumn{1}{c}{.787} & \multicolumn{1}{c}{.390} & \multicolumn{1}{c}{.445} & \multicolumn{1}{c}{} & \multicolumn{1}{c}{0.990} & \multicolumn{1}{c}{.957} & \multicolumn{1}{c}{.336} & \multicolumn{1}{c}{.801} & \multicolumn{1}{c}{.115} \\
\multicolumn{1}{c}{0.999} & \multicolumn{1}{c}{.973} & \multicolumn{1}{c}{.719} & \multicolumn{1}{c}{.797} & \multicolumn{1}{c}{.485} & \multicolumn{1}{c}{.514} & \multicolumn{1}{c}{} & \multicolumn{1}{c}{0.999} & \multicolumn{1}{c}{.994} & \multicolumn{1}{c}{.652} & \multicolumn{1}{c}{.924} & \multicolumn{1}{c}{.406} \\
\multicolumn{1}{c}{1.000} & \multicolumn{1}{c}{.975} & \multicolumn{1}{c}{.729} & \multicolumn{1}{c}{.798} & \multicolumn{1}{c}{.501} & \multicolumn{1}{c}{.529} & \multicolumn{1}{c}{} & \multicolumn{1}{c}{1.000} & \multicolumn{1}{c}{.996} & \multicolumn{1}{c}{.701} & \multicolumn{1}{c}{.926} & \multicolumn{1}{c}{.491} \\
\cmidrule{1-6}\cmidrule{8-12}\multicolumn{6}{c}{\textit{$T=100$}}          &       & \multicolumn{5}{c}{\textit{$T=1000$}} \\
\multicolumn{1}{c}{0.200} & \multicolumn{1}{c}{.000} & \multicolumn{1}{c}{.000} & \multicolumn{1}{c}{.000} & \multicolumn{1}{c}{.001} & \multicolumn{1}{c}{.002} & \multicolumn{1}{c}{} & \multicolumn{1}{c}{0.200} & \multicolumn{1}{c}{.000} & \multicolumn{1}{c}{.000} & \multicolumn{1}{c}{.000} & \multicolumn{1}{c}{.001} \\
\multicolumn{1}{c}{0.500} & \multicolumn{1}{c}{.000} & \multicolumn{1}{c}{.000} & \multicolumn{1}{c}{.000} & \multicolumn{1}{c}{.001} & \multicolumn{1}{c}{.078} & \multicolumn{1}{c}{} & \multicolumn{1}{c}{0.500} & \multicolumn{1}{c}{.000} & \multicolumn{1}{c}{.000} & \multicolumn{1}{c}{.000} & \multicolumn{1}{c}{.001} \\
\multicolumn{1}{c}{0.800} & \multicolumn{1}{c}{.041} & \multicolumn{1}{c}{.011} & \multicolumn{1}{c}{.031} & \multicolumn{1}{c}{.003} & \multicolumn{1}{c}{.184} & \multicolumn{1}{c}{} & \multicolumn{1}{c}{0.800} & \multicolumn{1}{c}{.000} & \multicolumn{1}{c}{.000} & \multicolumn{1}{c}{.000} & \multicolumn{1}{c}{.001} \\
\multicolumn{1}{c}{0.900} & \multicolumn{1}{c}{.458} & \multicolumn{1}{c}{.125} & \multicolumn{1}{c}{.321} & \multicolumn{1}{c}{.032} & \multicolumn{1}{c}{.243} & \multicolumn{1}{c}{} & \multicolumn{1}{c}{0.900} & \multicolumn{1}{c}{.000} & \multicolumn{1}{c}{.000} & \multicolumn{1}{c}{.000} & \multicolumn{1}{c}{.001} \\
\multicolumn{1}{c}{0.990} & \multicolumn{1}{c}{.965} & \multicolumn{1}{c}{.608} & \multicolumn{1}{c}{.827} & \multicolumn{1}{c}{.332} & \multicolumn{1}{c}{.425} & \multicolumn{1}{c}{} & \multicolumn{1}{c}{0.990} & \multicolumn{1}{c}{.900} & \multicolumn{1}{c}{.131} & \multicolumn{1}{c}{.623} & \multicolumn{1}{c}{.034} \\
\multicolumn{1}{c}{0.999} & \multicolumn{1}{c}{.983} & \multicolumn{1}{c}{.701} & \multicolumn{1}{c}{.849} & \multicolumn{1}{c}{.474} & \multicolumn{1}{c}{.528} & \multicolumn{1}{c}{} & \multicolumn{1}{c}{0.999} & \multicolumn{1}{c}{.996} & \multicolumn{1}{c}{.618} & \multicolumn{1}{c}{.942} & \multicolumn{1}{c}{.355} \\
\multicolumn{1}{c}{1.000} & \multicolumn{1}{c}{.985} & \multicolumn{1}{c}{.714} & \multicolumn{1}{c}{.850} & \multicolumn{1}{c}{.495} & \multicolumn{1}{c}{.546} & \multicolumn{1}{c}{} & \multicolumn{1}{c}{1.000} & \multicolumn{1}{c}{.998} & \multicolumn{1}{c}{.697} & \multicolumn{1}{c}{.947} & \multicolumn{1}{c}{.488} \\
\cmidrule{1-6}\cmidrule{8-12}\multicolumn{6}{c}{\textit{$T=200$}}          &       & \multicolumn{5}{c}{\textit{$T=5000$}} \\
\multicolumn{1}{c}{0.200} & \multicolumn{1}{c}{.000} & \multicolumn{1}{c}{.000} & \multicolumn{1}{c}{.000} & \multicolumn{1}{c}{.001} & \multicolumn{1}{c}{.000} & \multicolumn{1}{c}{} & \multicolumn{1}{c}{0.200} & \multicolumn{1}{c}{.000} & \multicolumn{1}{c}{.000} & \multicolumn{1}{c}{.000} & \multicolumn{1}{c}{.001} \\
\multicolumn{1}{c}{0.500} & \multicolumn{1}{c}{.000} & \multicolumn{1}{c}{.000} & \multicolumn{1}{c}{.000} & \multicolumn{1}{c}{.001} & \multicolumn{1}{c}{.057} & \multicolumn{1}{c}{} & \multicolumn{1}{c}{0.500} & \multicolumn{1}{c}{.000} & \multicolumn{1}{c}{.000} & \multicolumn{1}{c}{.000} & \multicolumn{1}{c}{.001} \\
\multicolumn{1}{c}{0.800} & \multicolumn{1}{c}{.000} & \multicolumn{1}{c}{.000} & \multicolumn{1}{c}{.000} & \multicolumn{1}{c}{.001} & \multicolumn{1}{c}{.138} & \multicolumn{1}{c}{} & \multicolumn{1}{c}{0.800} & \multicolumn{1}{c}{.000} & \multicolumn{1}{c}{.000} & \multicolumn{1}{c}{.000} & \multicolumn{1}{c}{.001} \\
\multicolumn{1}{c}{0.900} & \multicolumn{1}{c}{.084} & \multicolumn{1}{c}{.012} & \multicolumn{1}{c}{.049} & \multicolumn{1}{c}{.004} & \multicolumn{1}{c}{.182} & \multicolumn{1}{c}{} & \multicolumn{1}{c}{0.900} & \multicolumn{1}{c}{.000} & \multicolumn{1}{c}{.000} & \multicolumn{1}{c}{.000} & \multicolumn{1}{c}{.001} \\
\multicolumn{1}{c}{0.990} & \multicolumn{1}{c}{.968} & \multicolumn{1}{c}{.529} & \multicolumn{1}{c}{.844} & \multicolumn{1}{c}{.249} & \multicolumn{1}{c}{.389} & \multicolumn{1}{c}{} & \multicolumn{1}{c}{0.990} & \multicolumn{1}{c}{.001} & \multicolumn{1}{c}{.000} & \multicolumn{1}{c}{.000} & \multicolumn{1}{c}{.001} \\
\multicolumn{1}{c}{0.999} & \multicolumn{1}{c}{.990} & \multicolumn{1}{c}{.686} & \multicolumn{1}{c}{.888} & \multicolumn{1}{c}{.457} & \multicolumn{1}{c}{.534} & \multicolumn{1}{c}{} & \multicolumn{1}{c}{0.999} & \multicolumn{1}{c}{.996} & \multicolumn{1}{c}{.351} & \multicolumn{1}{c}{.932} & \multicolumn{1}{c}{.123} \\
\multicolumn{1}{c}{1.000} & \multicolumn{1}{c}{.992} & \multicolumn{1}{c}{.706} & \multicolumn{1}{c}{.889} & \multicolumn{1}{c}{.492} & \multicolumn{1}{c}{.562} & \multicolumn{1}{c}{} & \multicolumn{1}{c}{1.000} & \multicolumn{1}{c}{1.000} & \multicolumn{1}{c}{.697} & \multicolumn{1}{c}{.976} & \multicolumn{1}{c}{.487} \\
\cmidrule{1-6}\cmidrule{8-12}
\end{tabular}%
}
\caption{Simulation results. SVD, SVD* and BIC: unit root posterior probabilities (prior odds equal to one). DF: p-values of the Dickey-Fuller test. $\Pr_{\R \geq 1}$: posterior probabilities $\Prob\br{\R \geq 1|x}$.}
\label{tab:sim_p}
\end{table}

\section{Empirical Application}  \label{sec:empirical}
In our empirical application, we analyze Real Exchange Rates (RERs) time series for nine major currencies.
RERs are obtained by deflating nominal exchange rates by the relative price of domestic vs. foreign goods and services, thus reflecting the competitiveness of a country with respect to a reference basket. Common choices for deflation are the consumer price index (CPI), the producer price indices, or GDP-based deflators.
An increase in RER implies that exports become more expensive and imports turn cheaper, indicating a loss in trade competitiveness, for instance in response to an appreciation of the domestic currency, or in response to increased domestic inflation. 
We extract official monthly RERs (CPI deflated) time series distributed by the European Commission and available from the Statistical data warehouse of the European Central Bank\footnote{Data and its documentation are available at  \url{https://sdw.ecb.europa.eu/browse.do?node=9691113}.}, for the period between January 2010 and November 2020. This corresponds to 131 records for each of the nine\footnote{Australian dollar (AUD), Canadian dollar (CAD), Swiss franc (CHF), Chinese yuan (CNY), euro (EUR), British pound (GBP), Hong Kong dollar (HKD), Japanese yen (JPY),  US dollar (USD).} major currencies we consider in the analysis.

\begin{figure}[ht]
    \centering
    \includegraphics[trim={2.8cm 7.9cm 2.8cm 7.9cm},clip,scale=0.65]{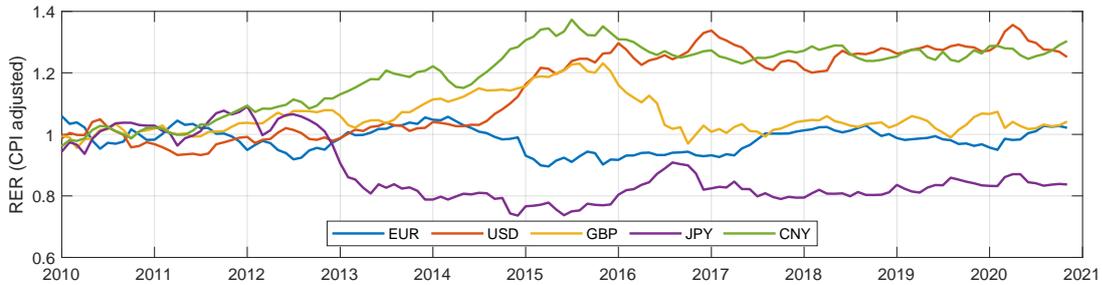}
    \caption{Real exchange rates for some selected currencies.}
    \label{fig:rates}
\end{figure}

Real exchange rates relate to the long-run equilibrium condition -- known as Purchasing Power Parity (PPP) -- which implies a steady long-term level and a constant unconditional mean for RERs series. The existence of a unit root would contradict this theory. There have been considerable efforts in empirically verifying the PPP theory and several papers discussed mid- and long- term departures from the expected RER stationarity, leading to a controversial debate. The essence is that conclusions based on empirical research strongly depend on the exact definition of equilibrium that one adopts, on the methods used to test it, on the underlying hypotheses on the time series, and on its length. This applies to unit root analyses as well, which often lead to opposite results on PPP's validity \citep[see e.g.][and the references therein]{macdonald1995long}. In this regard, the unit root analysis of \cite{10.2307/40440794} suggests that such equilibrium is expected to be generally observed over a time-span of at least 50 years. Non-short-lived disequilibrium periods in RER dynamics are thus common. At first sight, Figure \ref{fig:rates} seems to confirm such a tendency, suggesting a recent disequilibrium period for some of the major currencies. For instance, we observe an upward trend for the US dollar series and an apparent non-mean-reversing behavior for the Chinese yuan and Japanese yen, which gradually adjust towards new RER levels, suggesting the presence of unit roots.

Posterior probabilities of the unit root hypothesis in RERs are reported in Table \ref{tab:REER_prob}. Our results indicate ubiquitous evidence in favor of the unit root hypothesis for all the currencies. Bayesian results are aligned with the p-values from the Dickey-Fuller (DF) test and posterior probabilities from \citep{phillips1991criticize}. Indeed, higher posterior probabilities are both associated with higher p-values of the Dickey-Fuller statistics and higher posterior probabilities $\Pr_{\R \geq1} = \Pr\br{\R \geq 1|x}$ associated with the non-stationary alternative. Explanatory are the results for the Euro-zone where the .047 p-value of the DF test indicates a mid rejection of the unit root hypothesis. This is aligned with Figure \ref{fig:rates} where the EUR series indeed displays a much stationary behavior than any other series. Accordingly, BIC probabilities take their smallest value across all the countries analyzed, with the stationary hypothesis having .455 probability, much smaller than the average .880 observed for all the other currencies, where the DF p-value is on average .625. This is also aligned with the conclusion one would draw from the SVD* approach, where the highest evidence in favor of stationarity is again associated with the euro series. This is reasonable, as euro-zone countries have a significant presence in the basket for determining RER's CPI correction (19 out of 37 countries in the basket adopt euro). The observed greater plausibility of the stationary hypothesis is here not surprising but expected and coherent. Lastly, note that BIC probabilities appear to uniformly dominate SVD* ones, this could perhaps be explained in terms of the typical bias towards stationarity implied by the use of uniform priors.

\begin{table}[ht]
    \centering
        \scalebox{0.9}{
\begin{tabular}{ccccccc}
      & \multicolumn{2}{c}{$\log BF_{01}$} & \multicolumn{2}{c}{Prob.} &       &  \\
\cmidrule(lr){2-3}   \cmidrule(lr){4-5}  
Currency & SVD*  & BIC   & SVD*  & BIC   & DF    & $\Pr_{\R \geq1}$ \\
\midrule
AUD   & -0.333 & 1.076 & .418  & .746  & .090  & .288 \\
CAD   & 1.455 & 2.431 & .811  & .919  & .676  & .705 \\
CHF   & 0.996 & 2.282 & .730  & .907  & .442  & .569 \\
CNY   & 1.848 & 2.044 & .864  & .885  & .904  & .875 \\
EUR   & -1.194 & 0.180 & .232  & .545  & .047  & .161 \\
GBP   & 0.403 & 1.802 & .599  & .858  & .240  & .374 \\
HKD   & 1.812 & 2.120 & .860  & .893  & .883  & .857 \\
JPY   & 1.154 & 2.369 & .760  & .914  & .533  & .553 \\
USD   & 1.641 & 2.349 & .838  & .913  & .802  & .778 \\
\bottomrule
\end{tabular}
}
    \caption{SVD* and BIC: log- Bayes factors and their corresponding posterior probabilities for the unit root hypothesis (prior odds equal to one). DF: p-values of the Dickey-Fuller test. $\Pr_{\R \geq 1}$: posterior probabilities $\Pr\br{\R \geq 1|x}$.}
    \label{tab:REER_prob}
\end{table}

\section{Conclusion}  \label{sec:conclusion}
Unit root testing has historically been among the most active areas in econometric research. With classical frequentist methods following Dicker Fuller (DF) statistics, being broadly adopted and established, Bayesian methods did not gain much attention in empirical research and applications. However, research and debates on Bayesian unit root testing have been very active and sound. Indeed, the econometric research in the field pointed out a series of unique criticalities that turn the Bayesian approach for unit root inference particularly challenging.

On the other hand, the testing procedure based on the BIC approximation of the Bayes factor addressed in this paper appears to provide a simple and satisfactory method for Bayesian unit root testing. With such an approach, the integration problem involved in Bayes factors turns to a standard maximum likelihood estimation problem, and the Bayes factors have the very simple form of Eq. \eqref{eq:log_BIC}. Notably, (i) priors are not involved in the approximated form (though they determine the asymptotic error rate), (ii) the procedure smoothly scales to more complex time series models, and (iii) allows to target the exact point null unit root hypothesis. The ratio of the Bayes factor discriminates between the hypotheses according to the common interpretation scale of \citep{jeffreys1961theory}.

The simulation study confirms the validity of our proposed approach showing that, in this controlled setting, the BIC approximation, leads to decisions that are entirely coherent with the expectations, under a wide range of values for the autoregressive parameter, and sample sizes.
The posterior probabilities associated with the null are furthermore aligned with those of other Bayesian procedures and in accordance with the p-values from the DF test.
The same coherence between BIC, other Bayesian methods, and the frequentist DF test also arises from the the analysis of real exchange rates series. In particular, our BIC-based conclusion of non-stationarity matches the decisions one would draw based on the DF and SVD tests. The results are furthermore aligned with the posterior probabilities from \citep{phillips1991criticize}, in support of an apparent violation of the real exchange rate and purchasing power parity equilibrium in the last decade.

Recognizing that BIC is just a model Information Criterion (IC) among many others that perhaps reduce to BIC as special cases, it would be interesting to explore the use of such alternatives. 
Among these, generalized variants of BIC \citep[see e.g.][Ch. 9]{konishi2008information}, the FIC \citep{wei1992predictive}, and the ICs based on predictive distributions \citep{phillips1994posterior,phillips1996asymtotic} serve a broader model selection scope (than e.g. dampening the drawbacks associated to priors' selection in unit root testing) and rely on different theoretical and motivation bases. Generalized ICs are potentially superior for the generic purpose of model selection, however, they do not necessarily have a direct connection with well-identified Bayesian inference problems, though the use of some of them in unit root testing could now be explained by our results on BIC.
Perhaps a direction for related future works could be that of investigating to what extent ICs and model selection techniques have a clear link with certain problems in Bayesian inference. This would shed light on whether it is possible to rely on ICs and model selection methods as general tools for approximated Bayesian inference in situations where e.g. priors are of difficult specification. 

\section*{Acknowledgements}
This project has received funding from the European Union's Horizon 2020 research and innovation programme under the Marie Skłodowska-Curie grant agreement No. 890690.

\bibliography{Biblio}

\newpage
\appendix
\section{Appendices} 

\subsection{The model of Schotman and Van Dijk (1991b)} \label{sec:appendix_svd}

Consider the simplest autoregressive process of order one with zero-mean
\begin{equation}
    x_t = \rho x_{t-1} + u_t \, \text{.}
\end{equation}
Assume that (i) $x_0$ is a known constant, implying that we work conditionally on the initial observation, (ii) $u_t$ are independent and identically distributed (i.i.d.) normal random variables with mean zero and unknown variance $\S^2$, (iii) $\rho \in  \{ S,1 \}$; $S=\{ \rho|-1 < a \leq \rho < 1 \}$. We assume (iv) to observe a sample of $T$ observations on a time series $\{ x_t \}$. The Bayesian analysis is carried out via posterior odds. For the simple model here under consideration we have
\begin{equation}\label{eq:dijk_noconstant}
    K_1 = K_0 \frac{\int_0^\infty L\br{x|\rho = 1,\sigma,x_0} \pi\br{\sigma} \, d\sigma}{\int_S \int_0^\infty L\br{x|\rho,\sigma,x_0} \pi\br{\sigma}\pi\br{\rho} \, d \sigma d \rho } = \frac{\Prob\br{\rho = 1|x,x_0}}{\Prob\br{\rho \in S |x,x_0}} \, \text{.}
\end{equation}
where $K_0$ represents the prior odds ratio in favour of the hypothesis $\rho=1$, and $K_1$ the corresponding posterior odds ratio. The ratio between the integrals corresponds to the Bayes factor, $\pi\br{\sigma}$ and $\pi\br{\rho}$ represent the prior densities for $\sigma$ and $\rho \in S$, $L\br{x|\cdot}$ the likelihood function for the observed data $x$.

The prior odd $K_0$ expresses the relative weight of the null hypothesis against its stationary alternative such that the point $\rho=1$ is given the probability mass $\pi_0=K_0/\br{1+K_0}$, and analogously $K_1/\br{1+K_1}$ provides the posterior probability of the null hypothesis $\rho = 1$.

\cite{schotman1991bayesian} specify the marginal distribution of $\rho$ and $\sigma$ as
\begin{equation*}
    \Pr\br{\rho=1} = \pi_0 \, \text{,} \qquad 
    \Prob\br{\rho|\rho \in S} = \frac{1}{1-a} \, \text{,} \qquad 
    \Prob\br{\sigma} \propto \frac{1}{\sigma} \, \text{,}
\end{equation*}
that is, $\rho$ is taken uniform over $S$ and with probability mass $\pi_0$ on $\rho = 1$, with $\sigma $ and $\rho$ independent. 
Besides the fact that the density for $\rho$ depends only on the parameter $\alpha$ with great simplification of the integration problem in the denominator of Eq.\eqref{eq:dijk_noconstant}, the overall solution even in this simple setting is not obvious:
\begin{equation*}
    K_1 = \frac{\pi_0}{1-\pi_0} \frac{C_T^{-1}}{\br{T-1}^{1/2}}\br{\frac{\sigma^2_0}{\hat{\sigma}^2}}^{-\frac{T}{2}} \frac{1-a}{s_{\hat{\rho}}}  \left[ F\br{\frac{1-\hat{\rho}}{s_{\hat{\rho}}};T-1} -F\br{\frac{a-\hat{\rho}}{s_{\hat{\rho}}};T-1} \right]^{-1} \, \text{.}
\end{equation*}
$\hat{\rho}$ is the OLS estimator of $\rho$, $s^2_{\hat{\rho}}$ the squared OLS standard error of
$\hat{\rho}$, $\hat{\sigma}^2$ the estimated variance of the residuals, $\sigma^2_0$ the variance of
the first differences in $x$, $F\br{\cdot,\nu}$ the cumulative density of the t-distribution with $\nu$ 
degrees of freedom, $C_T = \Gamma\br{\br{T-1}/2}\Gamma\br{1/2}/\Gamma\br{T/2}$ a constant and $\pi_0/\br{1-\pi_0}$ the prior odds ratio $K_0$.

By choosing a prior equal balance between the stationary and random walk hypothesis $\pi_0 = 1/2$ \cite{schotman1991bayesian} recover a feasible operational test procedure based on an empirical determination of the lower bound $a$, 
$$
a^* = \hat{\R} + s_{\hat{\R}}F^{-1}\br{\alpha F\br{-\hat{\tau}}} \, \text{,}
$$
where $\tau = \frac{1-\R}{s_{\hat{\R}}}$ is the Dickey-Fuller statistics, and $0<\alpha<1$ a constant (typically between 0.001 and 0.1) such that the posterior contains $1-\alpha$ of its probability mass in $\left[a^*, 1\right)$. Therefore Bayes factors turn to be entirely functions of the data, and are computed as:
$$
K_1 = \frac{C_T^{-1}}{\br{T-1}^{1/2}}   \br{\frac{\S_0^2}{\hat{\S}_0^2}}^{-\frac{T}{2}}\br{ \frac{-\hat{\tau} - F^{-1}\br{\alpha F\br{-\hat{\tau}}}}{F\br{-\hat{\tau}}}}
\, \text{.}
$$

\subsection{The model of Phillips (1991b)} \label{sec:appendix_phi}

\cite{phillips1991criticize} adopts the information matrix prior from \citep{jeffreys1946invariant}. In particular, for a generic family of densities with parameter $\theta = \br{\R,\S}$ and information matrix $i$, the uninformative Jeffreys' prior he considers it is defined as $\pi\br{\theta} \propto \vert i \vert ^{\frac{1}{2}}$.

For the AR(1) model $x_t = \rho x_{t-1} + u_t$ with $u_t$ i.i.d. zero-mean normal with variance $\S^2$, and initial value $x_0$, the above prior becomes
$$
\pi\br{\rho,\sigma} \propto \frac{1}{\sigma}I_{\R}^\frac{1}{2} \,\text{,}
$$
where the continuous function $I_{\R}$, for $-\infty < \rho < +\infty$  and sample size $T$, is defined as:
\[
    I_{\R} = 
\begin{dcases}
    \frac{T}{1-\rho^2} - \frac{1}{1-\rho^2}\frac{1-\rho^{2T}}{1-\rho^2}+ \br{\frac{x_0}{\sigma}}^2 \frac{1-\rho^{2T}}{1-\rho^2}& \text{if }  \rho \neq 1 \text{,}\\
    \frac{T\br{T-1}}{2}+T\br{\frac{x_0}{\sigma}}^2              & \text{if }  \rho = 1 \text{.}
\end{dcases}
\]\\
This choice of the prior achieves tighter confidence sets for large values of $\vert \rho \vert$, is invariant to transformations of the parameter and enjoys other desirable properties. The prior depends on $x_0$ and its information grows with the sample size $T$ at a geometric rate when $\rho > 1$.
Under the Gaussian likelihood for the observed sample $x$ and the above prior, the posterior distribution of $\rho$ reads
\begin{equation} \label{eq:posterior_philips}
   \pi\br{\rho|x} \propto \alpha_0^\frac{1}{2} \left [R + \br{\rho - \hat{\rho}}^2 Q \right]^{-\frac{T}{2}} \text{.} 
\end{equation}
The practical contribution to the posterior from the choice of the prior is embedded in the factor $\alpha_0$:
\[
    \alpha_0 = 
\begin{dcases}
    \frac{T}{1-\rho^2} - \frac{1}{1-\rho^2}\frac{1-\rho^{2T}}{1-\rho^2} & \text{if }  \rho \neq 1 \text{,}\\
    \frac{T\br{T-1}}{2}              & \text{if }  \rho = 1 \text{.}
\end{dcases}
\]\\
On the other hand, the sum of squared residuals $R$, the quantity $Q = \sum x_{t-1}^2$ and the OLS estimate of the autoregressive coefficient $\hat{\rho}$ are entirely depend on the data and the model choice. The posterior in Eq.\eqref{eq:posterior_philips} is generally less susceptible of the downward bias than the posterior based on a flat prior, is not symmetric around $\hat{\rho}$, and depending on the values of the data-dependent quantities described above may have a second mode for $\rho >1$.

\subsection{Simulation study: complement} \label{sec:appendix_lbf}

\begin{table}[h!]
    \centering
    \scalebox{0.8}{
\begin{tabular}{crrrrrrrrr}
& \multicolumn{3}{c}{$T=50$} & \multicolumn{3}{c}{$T=100$} & \multicolumn{3}{c}{$T=200$} \\
\cmidrule(lr){2-4} \cmidrule(lr){5-7} \cmidrule(lr){8-10}
$\rho$ & \multicolumn{1}{c}{SVD} & \multicolumn{1}{c}{SVD*} & \multicolumn{1}{c}{BIC} & \multicolumn{1}{c}{SVD} & \multicolumn{1}{c}{SVD*} & \multicolumn{1}{c}{BIC} & \multicolumn{1}{c}{SVD} & \multicolumn{1}{c}{SVD*} & \multicolumn{1}{c}{BIC} \\
\midrule
0.200 & -11.35 & -11.60 & -11.13 & -23.81 & -24.25 & -23.60 & -48.97 & -49.54 & -48.77 \\
0.500 & -5.72 & -6.25 & -5.62 & -12.61 & -13.36 & -12.52 & -26.59 & -27.51 & -26.51 \\
0.800 & -0.88 & -1.95 & -1.15 & -3.16 & -4.52 & -3.43 & -8.08 & -9.66 & -8.36 \\
0.900 & 0.78  & -0.66 & 0.18  & -0.17 & -1.95 & -0.75 & -2.39 & -4.43 & -2.97 \\
0.990 & 3.06  & 0.68  & 1.31  & 3.31  & 0.44  & 1.56  & 3.42  & 0.12  & 1.68 \\
0.999 & 3.60  & 0.94  & 1.37  & 4.08  & 0.85  & 1.73  & 4.58  & 0.78  & 2.07 \\
1.000 & 3.68  & 0.99  & 1.38  & 4.19  & 0.91  & 1.73  & 4.76  & 0.88  & 2.08 \\
\midrule
& \multicolumn{3}{c}{$T=500$} & \multicolumn{3}{c}{$T=2000$} & \multicolumn{3}{c}{$T=5000$} \\
\cmidrule(lr){2-4} \cmidrule(lr){5-7} \cmidrule(lr){8-10}
$\rho$ & SVD   & SVD*  & BIC   & SVD   & SVD*  & BIC   & SVD   & SVD*  & BIC \\
\midrule
0.200 & -125.05 & -125.74 & -124.85 & -252.40 & -253.15 & -252.20 & -1273.42 & -1274.25 & -1273.21 \\
0.500 & -69.28 & -70.35 & -69.20 & -140.86 & -142.01 & -140.78 & -715.38 & -716.66 & -715.30 \\
0.800 & -23.43 & -25.23 & -23.71 & -49.41 & -51.34 & -49.69 & -259.36 & -261.48 & -259.65 \\
0.900 & -9.61 & -11.93 & -10.20 & -22.09 & -24.57 & -22.69 & -123.80 & -126.54 & -124.41 \\
0.990 & 3.10  & -0.68 & 1.39  & 2.20  & -1.89 & 0.50  & -7.01 & -11.63 & -8.74 \\
0.999 & 5.17  & 0.63  & 2.50  & 5.56  & 0.48  & 2.78  & 5.45  & -0.62 & 2.62 \\
1.000 & 5.57  & 0.85  & 2.53  & 6.21  & 0.84  & 2.88  & 7.78  & 0.83  & 3.69 \\
\bottomrule
\end{tabular}
}
\caption{Simulation results, log- Bayes factors. Refer to Section \ref{sec:simulation} for a discussion.}
\label{tab:sim_bf}
\end{table} 

\end{document}